\documentclass[aps,twocolumn,prd]{revtex4}

\usepackage{url}
\usepackage{latexsym}
\usepackage{epsfig}
\usepackage{scrextend}
\usepackage{lipsum}   

\usepackage{graphicx}

\usepackage{natbib}

\usepackage{xcolor}
\usepackage{amssymb}
\usepackage{amsmath}
\usepackage{graphicx}

 \definecolor{blue}{RGB}{7,80,201}  
 \definecolor{red}{RGB}{200,20,1}

\def\be{\begin{equation}}
\def\ee{\end{equation}}
\def\bea{\begin{eqnarray}}
\def\eea{\end{eqnarray}}
\def\ba{\begin{array}} 
\def\ea{\end{array}}
\def\bc{\begin{center}}
\def\ec{\end{center}}

\def\ghost#1{}

\def\simge{\mathrel{%
   \rlap{\raise 0.511ex \hbox{$>$}}{\lower 0.511ex \hbox{$\sim$}}}}
\def\simle{\mathrel{
   \rlap{\raise 0.511ex \hbox{$<$}}{\lower 0.511ex \hbox{$\sim$}}}}

\def\dis{\displaystyle}

\newcommand{\footremember}[2]{%
    \footnote{#2}
    \newcounter{#1}
    \setcounter{#1}{\value{footnote}}%
}

\begin{document}

\title
{\boldmath The light $U$ boson \vspace{3mm}as the mediator of a new force, \vspace{6mm}\hbox{coupled to a combination of $\,Q,\ B,\ L\,$ and dark matter} }

\author{{\sc P}{\small ierre} {\sc Fayet}
\vspace{3mm} \\ \small }

\affiliation{Laboratoire de Physique Th\'eorique de l'\'Ecole Normale Sup\'erieure\vspace{.3mm}\\ 24 rue Lhomond, 75231 Paris cedex 05, France 
\hspace{-.5mm}\footremember{a}{\vspace*{0mm} \  \em CNRS UMR 8549, \vspace{0mm}ENS, PSL Research University \,\em \& \em UPMC, \vspace{.8mm}Sorbonne Universit\'es}
\vspace{.8mm}\\
\hbox{and Centre de Physique Th\'eorique, \'Ecole polytechnique, 91128 Palai\-seau cedex, France}
\hspace{-.5mm}\footremember{b}{ \ \ \,\em CNRS UMR 7644, Universit\' e Paris-Saclay}
\vspace{2mm}\
\vspace{2mm}}

\begin{abstract}
\vspace{.2mm}
\hspace{0mm}A new light gauge boson $U$  may have both vector and axial couplings.
In a large class of theories however, the new $U(1)$ current $J^\mu_F$ naturally combines with the weak neutral current $J^\mu_{Z_{\rm sm}}$, both parity-viola\-ting, into a vectorial current $J^\mu_U$, combination of   the $B,\,L$ and electromagnetic currents with a possible dark matter current.

\vspace{1.2mm}

\hspace{0mm}$U^\mu $ may be expressed equivalently as 
\vspace{-.4mm}
$\,\cos\xi \,C^\mu\! + \sin \xi \,Z^\mu_{\rm \,sm}\,$ (``mixing with the $Z$'')  or  $\,(1/\!\cos\chi) \,\hat C^\mu+$ $\tan\chi \,A^\mu\,$
(``mixing with the photon''), with $\hat C$ coupled to $B,\,L$ and dark matter.
The $U$ boson may be viewed  as a generalized dark photon, coupled to SM particles through 
$\,Q_U\!=Q+\lambda_B B+\lambda_i L_i$, with strength  $\,g''\!\cos\xi\cos^2\theta=e\tan\chi$\,.
``Kinetic mixing'' terms, gauge invariant or not, simply corres\-pond to a description in a non-orthogonal field basis (rather than to a new physical effect), with  the dark photon in general also coupled to $B$ and $L$\,.

\vspace{1.2mm}
\hspace{0mm}In a grand-unified theory 
$Q_U^{\rm gut}\!=Q-2\,(B\!-\!L)$ at the GUT scale for SM particles, invariant under the $SU(4)$ electrostrong symmetry group, with a non-vanishing 
$\epsilon \!= \tan\chi$ already present at the GUT scale,  leading to $Q_U\simeq  \,Q-1.64\, (B-L)\,$ at low energy.
This also applies, for a very light or massless $U$ boson, to a new long-range force, which could show up through apparent violations of the Equivalence Principle.

\vspace{3mm}

\vspace{3.5mm}

\hfill Preprint LPTENS 16/07 \ \ \ \ \ November 15,  2016

\vspace{8mm}

\end{abstract}

\maketitle

\section{Introduction}

\vspace{-.5mm}

    The possible existence of a  light neutral spin-1 boson with a small gauge coupling,  in the  $\sim$ MeV to hundred MeVs  mass range and decaying most notably into $e^+ e^-$ pairs, 
    has been studied for a long time \cite{npb81,plb80,plb86}.  
It is generally expected not to have a significant effect on neutral current phenomenology at higher $q^2$, as compared to a heavy $Z'$,
  but  it could  affect the anomalous magnetic moments of the muon or electron, parity-violation effects in atomic physics, or be produced in various decays and beam dump experiments, etc..
    
    \vspace{1.5mm}
    
      In an electroweak theory the new boson, referred to as the $U$ boson, can mix with the $Z$ and with  the photon through a $3\times 3$ matrix,  borrowing features from both kinds of particles. 
  The $Z$ weak neutral current as well as the electromagnetic current, and also the $B$ and $L$ currents,  can contribute to the new current $J_U^\mu$, which has in general both vector and axial parts. This could however lead 
  to too-strong parity-violation effects \cite{plb05}, with axial couplings also enhancing the cross-sections for longitudinally-polarized $U$ bosons, produced  much like light pseudoscalars \cite{npb81}.

    \vspace{1.5mm}

 In a large class of spontaneously broken gauge  theories however, the $U$ couplings to quarks and leptons are {\it naturally vectorial in the small mass limit}.  
The $U$ current, obtained from a mixing of the extra-$U(1)$ current $J^\mu_F$ with  the $Z$ current, ultimately  involves a combination of the baryonic, leptonic (or $B-L$) and {electromagnetic} currents  \cite{plb89}. This includes and generalizes, within the framework of an extended electroweak or grand-unified theory,  the  very specific case of a ``dark photon'' coupled to electric charges, which has focused much of the experimental attention.
 This also applies to an extremely light or massless $U$ boson inducing a new long-range force, extremely weak, which could lead to apparent deviations from the Equivalence Principle \cite{plb86,plb89,newint}.

  \vspace{1.5mm}

The $U$ boson may also couple to dark matter, and can mediate sufficient annihilations  through {\it stronger-than-weak\,} interactions so as  to allow for dark matter particles to be light \cite{npb04,prd04}.
 It may provide a possible explanation for the apparent discrepancy between the expected and measured values of  $\,g_\mu\!-2\,$ \cite{prd07,pos,pdg},  and be at the origin of many interesting effects.

  \vspace{-1.5mm}

\section{\boldmath A vectorially-coupled $U$ boson}

\vspace{-.5mm}

In the presence of weak-interactions one may expect the $U$ couplings to be parity-violating. 
Still this one may be vectorially coupled to quarks and leptons, as in the specific case of a ``dark photon'', but allowing for more general situations. This occurs in a large class of models in which the extended electroweak symmetry is broken by a single doublet; or by two  (or more) as in supersymmetric theories, but with the same gauge quantum numbers. The symmetry breaking may then be viewed as induced by a single active doublet $\varphi$, the others being ``inert''.

\vspace{-2mm}

\subsection{\boldmath $Q_U$ as a combination of $Q,\,B,\,L$ and $F_{\rm dark}$}
\vspace{-1mm}

We express the extended electroweak covariant derivative as 

\vspace{-6mm}
\be
\label{cov}
iD_\mu =\,i\partial_\mu \!
\underbrace{- \,g\, \hbox{\boldmath $T$}.\hbox{\boldmath $V$\!}_\mu-\,\frac{g'}{2}\,Y\,B_\mu}_{\hbox{\small (possibly grand-unified)}}
-\,\frac{g''}{2}\,F\,C_\mu\,,
\ee
with  the extra $U(1)_F$ gauge symmetry commuting with $SU(3)_C\times SU(2)\times U(1)_Y$, or $SU(5)$ in a  grand-unified theory.
 The Lagrangian density, expressed as usual in an orthonormal field basis, includes the couplings of the gauge fields with the corresponding currents,
 \vspace{-1.5mm}
 \be
 \label{lcur}
 \ba{ccl}
\hspace{-1mm}{\cal L}\!&=&\!-\,\hbox{\small$\dis \frac{1}{4}$}\, \left( \hbox{\boldmath $V\!$}^{\mu\nu} \hbox{\boldmath $V\!$}_{\mu\nu}+ 
B^{\mu\nu}B_{\mu\nu}+ C^{\mu\nu}C_{\mu\nu}\right)
\vspace{1mm}\\
\!&&\!
\,-\ \hbox{\boldmath ${\cal J}$}^\mu. \,\hbox{\boldmath $V\!$}_\mu\,- {\cal J}_{\,Y}^\mu\,B_\mu\, -{\cal J}_{\,F}^\mu\,C_\mu + \,...\ \,.
\ea
\ee

\vspace{1mm}
When quarks and leptons acquire their masses from a single electroweak doublet as in the standard model, or several but with the same gauge quantum numbers, the gauge invariance of their Yukawa couplings requires the new $U(1)_F$ quantum number $F$, and the corresponding  current
${\cal J}^\mu_F$, to be expressed as \cite{plb89}
\be
\label{f}
\ba{ccc}
\ \ \ \ \ \ \ \ \ F&=&\!\alpha_B \,B\,+\,\beta_i \,L_i\,+\,\gamma \,Y\,+\,F_d\,,
\vspace{2mm}\\
{\cal J}^\mu_{F}\,=\,\hbox{\small $\dis \frac{g''}{2}$} \, J^\mu_F &=& \alpha_B\, {\cal J}^\mu_B+\beta_i\, {\cal J}^\mu_{L_i}+\gamma\, {\cal J}^\mu_Y+\,{\cal J}^\mu_{\,d}\,.
\ea
\ee
The new $U(1)$ current is naturally expressed as a linear combination of the $B$ and $L$ currents with the weak hypercharge current \footnote{
Two doublets, such as $h_1$ and $h_2^c$ in supersymmetric theories, both with $Y\!=-1$, also allow for a $U(1)_A$ symmetry rotating differentially the two doublets and acting axially on quarks and leptons. Its generator $F_A$ may provide an extra contribution to $F$ and ${\cal J}^\mu_F$ in (\ref{cov}-\ref{f}), the resulting current ${\cal J}^\mu_U$ including an axial part. A light $U$ boson is then produced very much like a light
pseudoscalar $a$, with the $U(1)_F$ symmetry broken at a scale larger than electroweak  through a large singlet v.e.v.\,. This ensures that this effective pseudoscalar $a$ has reduced interactions, very much as for an invisible axion~\cite{npb81,plb80}.}, 
and a possible dark-matter or extra spin-0 singlet contribution associated with a ``hidden sector''.  The simultaneous appearance of $B,\,L$ and $Y$ in 
(\ref{f}) is actually required in the framework of grand-unifica\-tion, to ensure that $U(1)_F$ commutes with the non-abelian grand-unification gauge group \cite{plb86,plb89}.
\vspace{2mm}

For $\gamma\!=\!0$ the new inter\-action,  coupled to a linear combination of the baryon and lepton numbers, is  simple to study in terms of the mass and couplings of the new boson, then unmixed with the $Z$ and the photon. For the theory to be anomaly-free the $U$ current may be taken as the $B-L$ current (in the presence of $\nu_R$ fields),
possibly combined with the $\,L_i-L_j$ and  dark matter currents. The new force may be of infinite or finite range, and may also act on dark matter particles.

\vspace{2mm}
 
We now concentrate on the more elaborate situation of  a $U(1)_F$ gauge interaction of $\,C^\mu$ in (\ref{cov}-\ref{f}) involving  the weak hypercharge generator $Y$, allowing to normalize $g''$ and $F$ so that $\gamma=1$. We may again use $B-L$ and $L_i-L_j$ in expression (\ref{f}) of $F$ for the theory
\vspace{-.4mm}
 to be anomaly-free, including $\nu_R$ fields.
The v.e.v. $v/\sqrt 2\simeq 174$ GeV of the doublet $\varphi$, with $Y\!=\!1\,$, breaks the $\,SU(3)_C\times SU(2) \times U(1)_Y\times U(1)_F$ gauge sym\-metry to  $SU(3)_C\times U(1)_{\rm QED}\times U(1)_U$, 
with 
$m_W=gv/2$\,. The three neutral fields  $\,W_3,\,B$ and $C$ are mixed into
the massless photon field $A$, the massive $Z$ field and a new neutral field $U$, still massless at this stage. They are given by \cite{plb89}
\vspace{.5mm}
\be
\label{33}
\framebox [7.6cm]{\rule[-1.7cm]{0cm}{3.6cm} $ \dis
\left\{\, \ba{ccl}
A\!&=&\!\hbox{$\dis \frac{g' \,W_3+g\,B}{\sqrt{g^2+g'^2}}$}\,=\,\sin\theta \, W_3+\cos\theta \, B\,,
\vspace{2mm}\\
Z\!&=&\!  \hbox{$\dis \frac{g \,W_3-g'B-g''C}{\sqrt{g^2+g'^2+g''^2}}$} \ ,
\vspace{2mm}\\
U\!&=&\!  \hbox{$\dis \frac{g''(g \,W_3-g'B)+(g^2+g'^2)\,C}{\sqrt{g^2+g'^2}\,\sqrt{g^2+g'^2+g''^2}}$}\ .
\ea
\right.
$}
\ee

\vspace{.5mm}

\begin{figure}[t]\centering
	\includegraphics[scale=0.6]{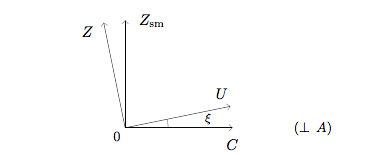}

%
%
%
%
%

\vspace{-1mm}
\caption{{\bf \em \boldmath ``Mixing with the $Z$''}, 
\vspace{.5mm}\em in the plane orthogonal to the photon field $A^\mu$.
\vspace{.3mm}
$U^\mu\!$ is a (small) mixing of  $\,C^\mu\!$ (coupled to $F=Y +\alpha_B B +\beta_i L_i+ F_d$)  with  $Z_{\rm sm}^\mu $,
\vspace{-.2mm}
leading to  ${\cal J}^\mu_U$
as in (\ref{curr},\ref{curr2}), with $\,\tan\xi = g''/\sqrt{g^2+g'^2}\,$.
\vspace{1mm}}
\label{fig:1}
\end{figure}

\noindent
The photon field has its usual SM expression in terms of $\tan \theta\!=g'/g$. $Z$ and $U$ are obtained by rotating the standard
$\,Z_{\rm sm}=\cos\theta \ W_3- \sin\theta \,B$\, and the $U(1)_F$ gauge field $C$ in the plane orthogonal to $A$ as represented in Fig.\,\ref{fig:1},
according to 
\be
\label{22}
\left\{\, \ba{ccl}
Z\!&=&\!  \cos\xi \ Z_{\rm sm} - \sin\xi \ C\,,
\vspace{2mm}\\
U\!&=&\!  \sin\xi \ Z_{\rm sm} + \cos\xi \ C\,,
\ea
\right.
\ee
\vspace{-3mm}

\noindent
with
\vspace{-3mm}
\be
\label{xi}
\tan\xi\,=\hbox{$\dis \frac{g''}{\sqrt{g^2+g'^2}}$}\,= \,\hbox{$\dis \frac{g''}{g_Z}$}\  .
\ee
This leads to the $3\times 3$ orthogonal mixing matrix in (\ref{33}).

\vspace{2mm}

The $U$  field is still massless at this stage, 
and coupled to a conserved current.
This one  is obtained, in the small $m_U$ limit in which $U$ is almost exactly given by (\ref{33}),  as a combination of the $U(1)_F$ current $J^\mu_F$ 
\vspace{-.5mm}
with the standard weak neutral current $J^\mu_{Z_{\rm sm}}$, 
using  $\,\sin\xi\, \sqrt{g^2+g'^2}=g'' \cos\xi\,$ from (\ref{xi}). Including as in (\ref{lcur},\ref{f}) the coupling constants $\,g,\,g'/2,\,g''/2\,$ within  the currents $\,{\cal J}^\mu $ we have
\bea
{\cal J}_U^\mu\!\!&=&\! \cos\xi\, \underbrace{(\,\frac{g''}{2} \,J_F^\mu\,)}_{\hbox{\small${{\cal J}_F^\mu}$}} \,+\,\sin\xi \, \underbrace{\sqrt{g^2+g'^2}\,(J_3^\mu\!-\sin^2\theta\,J^\mu_{\rm em})}_{\hbox{\small${\cal J}_{Z_{\rm sm}}^\mu$}}
\nonumber
\eea

\vspace*{-5mm}
\be
\label{curr}
\hspace*{-5mm}
\ba{ccc}
\hspace{4mm}&=&\hspace*{-2mm}g'' \cos\xi \ \left[\ \frac{1}{2}\ J^\mu_Y+\hbox{\small$\dis \frac{1}{2}$}\ (\alpha_B \,J^\mu_{B}+\beta_i\,J^\mu_{L_i}+ J_d^\mu)\right.
\ 
\vspace{2mm}\\  
&& \hspace{39mm} \left. + \, (J_3^\mu\!-\sin^2\theta\,J^\mu_{\rm em})\, \right]\,.\!\!\!
\ea
\ee

\vspace{2mm}

With $Y/2$ (from $J^\mu_F$) and  $T_3$ (from $J^\mu_{Z_{\rm sm}}$)
\vspace{-.3mm}
combining into $\,Q=\frac{Y}{2}+T_3$ the axial part in the $U$ current for quarks and leptons disappears in the massless $U$ limit.
The resulting current is
\be
\label{curr2}
\framebox [8.6cm]{\rule[-.85cm]{0cm}{1.8cm} $ \dis
\ba{ccc}
{\cal J}_U^\mu\!\!&=&\!g'' \cos\xi\,\left(\cos^2\theta \ J^\mu_{\rm em}+
\hbox{\small$\dis \frac{1}{2}$}\,(\alpha_B J^\mu_{B}+\beta_i\,J^\mu_{L_i}\!+ J_d^\mu) \right)
\vspace{2mm}\\
&=&\! e\tan\chi \,\left(J^\mu_{\rm em}+
\hbox{\small$\dis \frac{1}{2\,\cos^2\theta}$}\, (\alpha_B \,J^\mu_{B}+\beta_i\,J^\mu_{L_i}\!+ J_d^\mu) \right).
\ea
$}
\ee

\vspace{1mm}

\noindent
It is associated with a conserved charge \cite{plb89}, normalized as
\be
\label{qu}
\framebox [6.8cm]{\rule[-.3cm]{0cm}{.8cm} $ \dis
Q_{U}\,= \,Q \,+  \, \hbox{\small$\dis \frac{1}{2 \,\cos^2\theta}$}\ (\alpha_B B+\beta_i L_i+F_d)\,,
$}
\ee
with coupling constant 
\be
\label{eps}
\ba{ccl}
g''\, \cos  \xi \,\cos^2 \theta \!&=&\! g''\,\hbox{ $\dis \sqrt{\frac{g^2\!+g'^2}{g^2\!+g'^2\!+g''^2}}\  \frac{g^2}{g^2+g'^2} $} 
\vspace{2mm}\\
&=&  \epsilon \,e\, =\, e\,\tan\chi \,\simeq\,g''\cos^2 \theta\,.
\ea
 \ee
 
 \vspace{2mm}

This coupling   is expressed
\vspace{-.2mm}
 in terms of the elementary charge $e=gg'/\!\sqrt{g^2+g'^2}\,$ as $\epsilon \,e$ with $\,\epsilon \!=\tan\chi$. The angle $\chi$ may be interpreted
 from another expression of 
 $U^\mu$, given in a non-orthogonal basis
 \vspace{-.5mm}
 as $(1/\cos\chi)\, \hat C^\mu\, + $ $\tan\chi \,A^\mu$ (cf.~eqs.\,(\ref{33bis},\ref{rotchi},\ref{ac}) and Fig.~\ref{fig:3} in Sec.\,\ref{sec:5}).
$\,\hat C^\mu$ is  in general coupled, not only to dark matter and to a \hbox{spin-0} field $\sigma$ in the hidden sector
 \vspace{-.2mm}
 responsible for $m_U$
\cite{plb89}, 
 but also to $B$ and $L$ as well. $\hat C^\mu$ and $A^\mu$, both orthogonal to $Z^\mu$ in (\ref{33}), are themselves 
non-orthog\-onal, at an angle $\frac{\pi}{2}+\chi$ with $\tan\chi =\epsilon$ given by (\ref{eps}), very close to $\frac{\pi}{2}$ if the extra $U(1)$ coupling $g''$ is small.
 
  \vspace{-1mm}
  
 \subsection{\boldmath Relations between $U$ charges}
 
 \vspace{-1mm}
 
The $U$ charges in (\ref{qu}), expressed for the first generation  of quarks and leptons as a linear combination of $Q,\,B$ and $L_e$, satisfy the same additivity  relations
\be
\label{rel}
\ba{ccl}
Q_U(p)-Q_U(n)=Q_U(u)-Q_U(d) \hspace{19mm}
\vspace{2mm}\\
\hspace{17mm} =Q_U(\nu_e)-Q_U(e)=\,Q_U(W^+)=\,1\,.
\ea
\ee
They express the conservation of $Q_U$ in the limit of a massless $U$, associated with an unbroken symmetry $U(1)_U$, with  the $W^\pm$ carrying $\,\pm\,1$ unit of $Q_U$.
The corresponding vector couplings of the $U$ boson, expressed from (\ref{curr2}-\ref{eps})  as 
\be
\left\{\,
\ba{cclcccc}
f_{\nu_e}\!\!\!&=&\!\epsilon e \ Q_U(\nu_e)\,, &f_p\!\!&=&\!\!2f_u\!+\!f_d\,=\, \epsilon e \ Q_U(p)\,, 
\vspace{1mm}\\
f_e\!&=&\! \epsilon e \ Q_U(e)\,,&f_n\!\!&=&  f_u\!+\! 2f_d=\, \epsilon e \ Q_U(n)\,, 
\ea \right.
\ee 
must there\-fore verify
\vspace{-.5mm}
\be
\label{rel2}
f_n = \,f_p+f_e-f_{\nu_e} \,.
\ee
This relation expresses in particular that $Q_U$ remains conserved in the $\beta$ decay of the neutron, $\,n \to  p \ e \ \bar\nu_e$\,.

\vspace{2mm}

 The $U$  is constrained to interact very weakly with electrons, so that the extra contribution $\delta_U a_e$ to the electron anomaly be sufficiently small;  
and with protons, so that the $\pi^0\!\to\! \gamma\, U$ decay amplitude, proportional to $ f_p\!=2f_u\!+\!f_d$, be sufficiently small (cf.~Sec.\,\ref{sec:6}).
\,It should also interact sufficiently weakly with neutrinos so as to satisfy the constraint
$|f_{\nu_e} f_e|/m_U^2$ $\simle G_F$ from  low-$q^2\ \nu_e$-$e$ scattering, i.e.~\cite{npb04,plb05}
\be
\label{nu}
|f_{\nu_e} f_e|^{1/2}\, \simle \ 3\ 10^{-6} \ m_U\hbox{(MeV)}\,,
\ee
valid for $m_U$ larger than a few MeVs.
The additivity property (\ref{rel}-\ref{rel2})
implies  that if $f_e,\,  f_p$ and $f_{\nu_e}$ are all small, the coupling to the neutron $\,f_n\!= f_p+f_e- f_{\nu_e}$ is  expected to be small as well.

\vspace{2mm}
The $U(1)_F$ generator in (\ref{f}) may well involve $B$ and $L$ through their difference \hbox{$B-L$}, in view of an anomaly-free theory (including $\nu_R$'s), or of grand-unification as we shall see. $Q_U$ may then be expressed as 
\be
\label{newq20-1}
\framebox [6cm]{\rule[-.25cm]{0cm}{.7cm} $ \dis
Q_U = \,Q\,-\,\lambda\,(B-L)\,+\,Q_{U\rm dark}\,.
$}
\ee
More specifically with $\lambda\simeq 1$\,,
\be
Q_U  \simeq\,  Q-(B-L)\,+\,Q_{U\,\rm dark}\,,
\ee
would lead to smaller interactions with the proton and the electron, i.e.
\be
\hbox{\em small \ $ (f_p=- f_e)$, \ as compared to} \ \ (f_n= -\,f_\nu)\,,
\ee
very much as found  in \cite{plb86}, in the presence of axial couplings.
In a similar way $Q_U$ close to $\,Q-B +Q_{Ud}$  (or $Q-(B-3L_\tau)+Q_{Ud}$ in view of an anomaly-free theory) would lead to small  $f_p$ and $f_{\nu_e}$, with a larger 
$f_n\simeq f_e$, again in agreement with (\ref{rel2}).
\vspace{2mm}

Such relations, however, may be avoided in other situations, with  two spin-0 doublets at least, in which the $U$ current is not naturally vectorial and conserved  in the small $m_U$ limit, so that significant parity-violation effects may have to be expected. A  light $U$ in a longitudinal polarisation state may then be 
produced and interact significantly, much like the \hbox{spin-0} pseudoscalar $a$ associated with the spontaneous breaking of the global $U(1)_U$ \cite{npb81}. Both effects restrict significantly the possible size of  axial couplings (cf.~subsection~\ref{subsec:6a}).

\vspace{-1.5mm}

\subsection{\boldmath A new long-range \vspace{1mm}force, and \hbox{\ \ \ \ \ \ Equivalence Principle tests}}

\vspace{-1mm}

The  $U$ could stay massless, mediating a new long-range force acting {\it additively\,} on ordinary particles, proportionally to a linear combination of $B$ (as considered long ago by Lee and Yang \cite{ly}), with $L$ and $Q$. Or it  may acquire a mass if the $U(1)_U$ symmetry gets spontaneously broken. The fact that both $B$ and $L$ can be present  simultaneously in the expression of $Q_U$, and in combination with  the electric charge $Q$, allows for an extension to grand-unified theories. 
The fact that $Q$ may also appear alone  illustrates that the popular dark photon case is included as a specific case of this general analysis.

\vspace{2mm}

For ordinary matter  $Q_U$ appears as 
a combination of the numbers of protons, neutrons and electrons, or $Z,\,N$ and $Q$, i.e.~effectively $Z$ and $N$ only for ordinary neutral matter \cite{newint}.  More specifically the new force may act mostly on the number of neutrons $N$, as in the case of a ``protophobic''  $U$ boson 
for which $Q_U$ is close to $Q-(B-L)$  \cite{plb86}.

\vspace{2mm}
With $\,Q_U\!=Q-\lambda\,(B-L)+Q_{U\rm d}$ as in (\ref{newq20-1}) (and also in the absence of the $Q$ term, if $\,Y$ does not appear in  expression (\ref{f}) of $F$ so that the $U$ does not mix with the $Z$ and the photon), one has
\be
Q_U(p+e)=0\,,\ \ Q_U(n)=-\lambda\,,
\ee
so that
\vspace{-3mm}
\be
Q_U=-\,\lambda \,N\,,
\ee
for ordinary neutral matter. The interaction potential between two bodies of mass $m_i$ and number of neutrons $N_i$ is then given by
\be
\framebox [8.2cm]{\rule[-.3cm]{0cm}{.95cm} $ \dis
V(r)=\,-\,\frac{G_N\,m_1m_2}{r}\,+\,\frac{(\lambda\epsilon)^2\,e^2\,N_1N_2}{4\pi\epsilon_\circ\,r}\,
e^{-\,r/\!\hbox{$\frac{\hbar}{m_Uc}$}}\ .
$}
\ee
 The ratio between the repulsive $U$-exchange potential and the gravitational potential between two neutrons at a distance $r$ somewhat larger than $\,\hbar/m_Uc$\, is about
 \be
 \frac{V_U(r)}{V_g(r)}\,\simeq \,-\,(\lambda\epsilon)^2\,\alpha\,\left(\frac{m_{\hbox{\tiny \,Planck}}}{m_n}\right)^2 \simeq \,-\,1.23 \ 10^{36}\,(\lambda\epsilon)^2\,.
 \ee

\vspace{2.5mm}

For a massless or almost massless $U$ boson the new force could lead to apparent deviations from the Equivalence Principle \cite{plb89,newint},
constraining it  to be considerably weaker than gravitation, by $\approx 10^{-10}$ at least, corresponding typically to 
\vspace{-2.5mm}
\be
\lambda\epsilon\, \simle \,10^{-23}\,,
\ee
depending also on $m_U$ and $\,\lambda_U\!=\hbar/m_Uc\,$, \,so that the resulting violations of the Equivalence Principle be $\simle 10^{-13}$ \cite{adel1,adel2,adel3,adel4}. 

\vspace{2mm}

The {\small MICROSCOPE} experiment will soon test the validity of this principle at the $10^{-15}$ level \cite{micro}.
The additivity property of the new force induced by a \hbox{spin-1} $U$ boson, following from the linear expression (\ref{qu}) of $Q_U$ (as opposed to
an hypothetical coupling to mass, or strangeness, ...~), is also in contrast with the case of a spin-0 mediator  \cite{damour}, for which other contributions to the expression of the new force are generally expected.   This may allow for {\it a distinction between spin-1 and spin-0 mediators}, should a deviation from the Equivalence Principle be observed.

\vspace{-1mm}

\subsection{\boldmath Generating a small mass for the $U$ boson}
\vspace{-.5mm}

The $U$ boson can acquire a small mass from a neutral singlet $\sigma$ with $Y\!=0$, directly providing  $A=\sin\theta\,W_3+$ $\cos \theta \,B\,$  in (\ref{33}) as the massless photon field    \cite{plb89}. 
\vspace{-.2mm}
The   singlet v.e.v.\,\,\,$<\!\sigma\!>\ =w/\sqrt 2$ generates a mass term $m_C=$ $g'' F_\sigma w/2$, 
resulting in a small $U$ mass 
\be
\label{mu}
m_U\,\simeq \,m_C\,\cos\xi\ \simeq \,g'' F_\sigma w/2\,,
\ee
with $\sqrt 2\ \Re\,  \sigma$ a physical singlet BEH field, possibly (slightly) mixed with the standard one 
$\sqrt 2 \ \Re \,\varphi^0$ taken to describe the new 125 GeV spin-0 boson  \cite{beh,beh2}.
\vspace{2mm}

The massive $U$ field  differs very little from its expression in (\ref{33},\ref{22}), through a tiny change in $\xi$ (from $\xi_\circ$ to $\xi_\circ+\delta \xi$) inducing very small parity-violating contributions $\approx m_U^2/m_Z^2$ in the $U$ current $J^\mu_U$. To discuss these small $m_U$ corrections to the $U$ and $Z$ currents 
we observe that the theory is invariant under a simultaneous change of sign for $C^\mu$ and $g"$, 
acting as
\be
\hspace{7mm}C^\mu\to\,-\,C^\mu\,,\ \ g"\to\,-\,g"\,,
\ee

\vspace{-2mm}

\noindent
so that 
\vspace{-2mm}
\be
\ba{rcc}
(A,\,Z,\,U)\ \ &\to&\ (A,\,Z,\,-\,U)\,,
\vspace{3mm}\\
\hbox{with}\ \  (g",\xi,\chi,m_U) \!&\to&\!-\, (g",\xi,\chi,m_U)\,.
\ea
\ee

Corrections to ${\cal J}^\mu_{\,U}$ in (\ref{curr},\ref{curr2}), odd in $g"$ and vanish\-ing with  $m_U$,
are thus $\,\approx g"\,\cos\xi \ m_U^2/m_Z^2$ (rather than $g"\,\cos\xi $ $\,m_U/m_Z$), i.e.
\be
\approx \,e\,\tan\chi \ m_U^2/m_W^2\,.
\ee
The $Z$ current  differs also very little from its  standard SM expression with a small contribution from  $J^\mu_F$\,,  by terms $\approx m_U^2/m_Z^2$ as obtained from (\ref{33}-\ref{curr2}).

\vspace{-1mm}

\subsection{Special case of the dark photon}

\vspace{-1mm}

The special case for which $B$ and $L$ do not participate  in expression  (\ref{f}) of  the  $U(1)_F$ quantum number, simply reduced \vspace{-2mm} to
\be
\label{yfd}
F= Y +F_d\,,
\ee
provides in an electroweak theory a ``dark photon'', with  the $U$ coupled  to standard model particles proportionally to their electromagnetic current,  through 
\be
{\cal J}^\mu_U = \,\underbrace{g''\cos\xi \,\cos^2\theta}_{\hbox{\normalsize $\epsilon\,e$}} \ (J^\mu_{\rm em}+ \hbox{\small$\dis \frac{1}{2\,\cos^2\theta}$}\,J^\mu_d\,)\,.
\ee
\vspace{-1.8mm}

\noindent
The  coupling $g''\cos\xi \cos^2\theta$ 
may be expressed  as in (\ref{eps}) as $\epsilon \,e$ in terms of the elemen\-tary charge 
$\,e=gg'/\!\sqrt{g^2+g'^2}$, 
with
\be
\label{eps2}
\framebox [8cm]{\rule[-.35cm]{0cm}{.9cm} $ \dis
\epsilon\,=\, \tan\chi \,= \,\frac{g''}{g'}\ \frac{g}{\hbox{$\dis\sqrt{g^2+g'^2+g''^2}$}}\,\simeq \,\frac{g''}{g'}\,\cos\theta\,,
$}
\ee
in the small $m_U$ limit.

\vspace{2mm}

This  simple situation, a special case of the general one, has been obtained in an extended electroweak theory without ever referring to a largely fictitious ``kinetic mixing'' term. Such terms are  simply associated with a description in {\em a non-orthogonal field basis, independently of the fact that they are gauge invariant or not}. Adding them explicitly in an initial Lagrangian density (then under the restrictive condition that they must be gauge invariant \cite{galison,hol}) does not provide additional physical freedom. Indeed the notion of scalar product is not generalized by adding to its usual expression $xx'+yy'$ in orthonormal coordinates a non-diagonal ``mixing term'' $\epsilon \,(xy'+yx')\,$.
Discussing a theory in a non-orthogonal rather than in an orthonormal field basis has no effect on the results. 

\vspace{2mm}

The simple situation of a ``dark photon''  has been the focus of much experimental attention recently \cite{dark,dark2}.
But it appears excessively restrictive as compared to the general situation for a $U$ boson \cite{plb89}, excluding possible contributions from the $B$ and $L$  currents, that ought to be present in a grand-unified theory.

\vspace{-2mm}

\section{\boldmath The \vspace{1.5mm} $\,U$ current \hbox{in grand-unified theories}}

Indeed within $SU(5)$-type grand-unified theories \cite{gg,gqw},
 the weak-hyper\-charge $Y\!$, \,now a generator of $SU(5)$, is no longer abelian, while the $U(1)_F$ generator $F$
in (\ref{cov},\ref{f}) should commute with $SU(5)$.
It may then look like gauge invariance  prevents $Y$ from entering in the expression of the $U(1)_F$ quantum number 
$F$ for the visible particles,
 requiring $\,\epsilon =\tan\chi$ to vanish at tree-level, as commonly believed \cite{dark2}.

\vspace{2mm}

This is not true however, as $Y$ can contribute to $F$
through the $SU(5)$-invariant combination
 involving \hbox{$B-L\,$} \cite{plb89}, 
 
 \pagebreak
 
 \vspace*{-8mm}
\be
\framebox [7.5cm]{\rule[-1.35cm]{0cm}{2.9cm} $ \dis
\ba{l}
\ F\,=\, Y-\frac{5}{2}\,(B-L) 
\vspace{3mm}\\ 
 \hspace{7mm} = \ \left\{\!\! \ba{cl}  
-\frac{1}{2}:&\underline{\hbox{\bf 10}} \ \  \hbox{\small of} \  SU(5)\  \ \{u_L, d_L,\bar u_L,e^+_L\}\,,
\vspace{2mm}\\
\ \ \frac{3}{2} : &\ \underline {\bar {\bf 5}} \  \ \ \ \hbox{\small of} \  SU(5) \ \ \ \{\bar d_L, \nu_L,e^-_L\}\,,
\vspace{2mm}\\
\ \ 1 :&\ \underline {\bf 5}_H \  \ \hbox{\small of} \  SU(5) \ \ \ \hbox{including}\ \varphi\ .
  \ea\right.
  \ea
  $}
\ee
One also has $F\!=-\frac{5}{2}$ for possible $\bar \nu_L$  singlets describing right-handed neutrinos $\nu_R$, making the theory anomaly-free, and the $U(1)_F$ generator traceless.
All Yukawa couplings proportional to $\underline{\bar {\bf 5}}_H \,.\, \underline{\bar{\bf 5}}\,.\, \underline{\bf 10}\,$  
 and 
$\underline {\bf 5}_H\,.\, \underline{\bf 10}\,.\, \underline{\bf 10}\,$, responsible for  down-quarks and char\-ged-lepton masses and up-quark masses, respectively, have $F\!=\!0$ and are invariant under $SU(5)\times U(1)_F$.

\vspace{2mm}

The  $U(1)_F$ current, including its hidden-sector part  $J^\mu_d$, now reads
\vspace{-3mm}
\be
J_F^\mu\,=\,J_Y^\mu\,-\,\frac{5}{2}\,J^\mu_{B-L} +\,J^\mu_d\,.
\ee
By combining it with  $J^\mu_{Z_{\rm sm}}$ as in (\ref{curr},\ref{curr2}) we get the $U$ current
\vspace{-3mm}
\be
\ba{ccl}
{\cal J}_U^\mu
\!\!&=&\!\! 
g'' \cos\xi \left(\frac{1}{2}J^\mu_Y-\frac{5}{4}J^\mu_{B-L}+\frac{1}{2}J_d^\mu+ (J_3^\mu\!-\sin^2\theta\,J^\mu_{\rm em})\right) 
\vspace{3mm}\\
\!\!&=&\  g'' \cos\xi\,\left(\,\cos^2\theta \ J^\mu_{\rm em}-\frac{5}{4}\ J^\mu_{B-L}+\frac{1}{2} \ J_d^\mu\,\right)\,,
\vspace{-4mm}\\
\ea
\ee

\vspace{3mm}

\noindent
associated with the charge  \cite{plb89}\be
\label{qu2}
\framebox [7.2cm]{\rule[-.35cm]{0cm}{.85cm} $ \dis
Q_{U}\,= \,Q \,-  \, \hbox{\small$\dis \frac{5}{4 \,\cos^2\theta}$}\ (B-L)+ \hbox{\small$\dis \frac{1}{2 \,\cos^2\theta}$} \ F_d\,.
$}
\ee
The corresponding current $J^\mu_U$ involves a term proportional to the electromagnetic current $J^\mu_{\rm em}$, with the same coupling 
\be
g'' \cos  \xi \cos^2 \theta = \epsilon\, e = e\,\tan \chi
\ee
 as in (\ref{eps},\ref{eps2}), and 
a term proportional to the $B-L$ current, plus a dark matter current.

\vspace{1mm}

\section{\boldmath $\!Q_U$, \vspace{1.5mm}\,the electrostrong symmetry, and $\,B-L$}
\vspace{-.5mm}

\subsection{\boldmath $Q_U$ at the grand-unification scale}

\vspace{-1.5mm}

To understand better the origin and meaning of the new charge  we note that
$Q_U$ in (\ref{qu2}),
\vspace{-.3mm}
 if evaluated at the grand-unification scale with  $g'\!= g\sqrt{3/5} $ and $\sin^2\theta =$ $3/8$, would read
 \vspace{-1.5mm}
\be
\label{qugut0}
\hbox{\small at GUT scale:}\ \ \ \ Q_U^{\rm \,gut}=\,Q-2\,(B-L) + \frac{4}{5}\,F_d\,.
\ee
\vspace{-3.5mm}

\noindent
This $Q_U^{\rm \,gut}$ is  invariant under  the $SU(4)_{\rm es} \,(\sim O(6))$ {\em electrostrong} symmetry group unifying directly electromagnetism with strong interactions,  including $SU(3)_C\times U(1)_{\rm QED}$ within $SU(5)$, and commuting with $U(1)_U$.
Indeed, forgetting momentarily about the \,\underline{\boldmath $24$} adjoint v.e.v.~(or extra dimensions) responsible for the breaking of 
$SU(5)\ [\times U(1)_F]$ into $SU(3)_C\times SU(2)\times U(1)_Y\ [\times U(1)_F]$,  the v.e.v.~of the quintuplet ${\underline{\bf 5}}_H$ (including the electroweak doublet $\varphi$, with $F=Y=1$),  is responsible for the symmetry breaking
\vspace{-1mm}
 \be
SU(5) \times U(1)_F   \stackrel{\ba{c}\hbox{\normalsize$<{\underline{\bf 5}}_H\!>$}\vspace{1mm}\\ \ea}{\longrightarrow}\ SU(4)_{\rm es}\times U(1)_U\,.
 \ee 
It  leaves at this stage unbroken the $U(1)_U$ group generated by $Q_U^{\,\rm gut }$ in (\ref{qugut0}), commuting with 
 $SU(4)_{\rm es}$. This remaining $U(1)_U$ may then be broken 
 \vspace{-.3mm}
 by the singlet v.e.v. $<\!\sigma\!>\ =w/\sqrt 2$,
 generating a non-vanishing $m_U$ as in (\ref{mu}).
 
 \vspace{2mm} 

This $U$ charge, invariant at the GUT scale under the $SU(4)_{\rm es}$ electro\-strong symmetry group, is the same  for all components within 
$SU(4)_{\rm es}$ representations, i.e.~for quarks and leptons
\vspace{-3mm}
\be
\label{qugut1}
Q_U^{\rm \,gut}\,=\ \left\{\ \ba{ccc} 1\,:&\  \  \underline{\bar{\bf 4}}
& \ \ \left(\ba{c}\bar d \\ \,e^-\!\! \ea\right)_{\!L+R},
\vspace{2.5mm}\\
0\,:& \  \ \underline{\bf 6} &  \ \{u,\,\bar u\}_L\ ,
\vspace{2.5mm}\\
2\,: &\  \ \underline{\bf 1} & \  \nu_L \,(+\,\nu_R)\ ,
\ea\right.
\ee
with family indices omitted for simplicity. Dirac quark and lepton fields (excepted possibly for chiral neutrinos) are in vectorial representations of 
$SU(4)_{\rm es}\times U(1)_U$, with the $U$ boson interactions, as well as electrostrong interactions, invariant under $SU(4)_{\rm es}$, preserving parity.

\vspace{2mm}
For the 24+1 gauge bosons of $SU(5)\times U(1)_F$, which includes $SU(4)_{\rm es}\times U(1)_U$ as a subgroup,  one has
\vspace{1mm}
\be
\label{qugut2}
Q_U^{\rm \,gut}=\,\left\{\ \ba{ccc} 0\,:& \  \underline{\bf 16}
&\ \{ \hbox{gluons},\, \gamma, \,X^{\pm4/3}\}\,,
\vspace{2mm}\\
0\,: & \ \ \underline{\bf 1} &  Z\,,\ U\,,
\vspace{2mm}\\
1\,:&\  \ \underline{\bf 4}  &\  \left(\!\ba{c}Y^{-1/3} \\ W^+\, \ea\!\!\right) \,,
\vspace{2mm}\\
-1\,:&  
\ \ \underline{\bar{\bf 4}} &\  \left(\,\ba{c}Y^{1/3}  \\ W^-\, \ea\right) \,.
\ea\right.
\ee
$Q_U^{\rm gut}$
vanishes as required for 
the $\{u,\,\bar u\}$ self-conjugate sextets (the $U$ being ``$u$-phobic''  at the GUT scale), with
\vspace{-3mm}
\be
\label{qugut}
\ba{c}
\hbox{$Q_U^{\rm gut}= \,Q-2(B\!-\!L):$}\ \, 
\left\{\ba{cc}Q_U^{\,p}=\ \, Q_U^{\,d}\,=\,-\,1\,,  &\ Q_U^{\,\nu}=2,
\vspace{1.5mm}\\
   Q_U^{\,n}= 2\,Q_U^{\,d}=\,-\,2\,,&\ Q_U^{\,e}=1,
   \vspace{1.5mm}\\
   Q_U^{\,u}=0\,.
   \ea\right.
   \vspace{-3mm}\\
   \ea
\ee

\begin{figure}[t]\centering 
\hspace*{-5mm}	\includegraphics[scale=0.65]{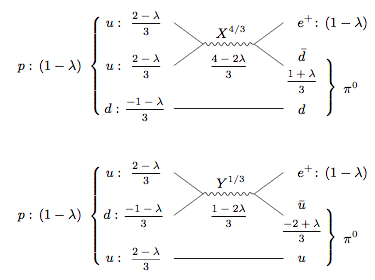}


\vspace{2mm}
\caption{\em  Two of the diagrams responsible for $\,p\to \pi^0 e^+\!$ in a $SU(5)\times U(1)_F$ theory, showing the values of 
 $\,Q_U\!=Q-\lambda(B-L) \,+Q_{\rm dark }$ and illustrating its conservation. $\lambda=2$ at the GUT scale (with $Q_U=0$ for $\,u$ and $X^{4/3}$), down to 1.64 at low energy. $\,Q_U\!= 1-\lambda\simeq -\,.64$ for both the proton and the positron emitted in its decay.}
\label{fig:gut}
\end{figure}

\vspace{-2mm}

\subsection{\boldmath $Q_U$ at low energy}

\vspace{-1.5mm}

Below the grand-unification scale expressions (\ref{qugut0}, \ref{qugut1}-\ref{qugut}) of $Q_U$ get modified as in (\ref{newq20-1}) according to
\be
\label{newq20}
Q_U = Q\,-\,\lambda\,(B-L)\,+\,Q_{U\,\rm dark}\,,
\ee
down to (\ref{qu2}),
\be
\label{newq3}
Q_U \simeq \,Q\,-\,1.64\,(B-L)\,=\left\{\! \ba{ll}
p: \ -\,.64\,;&\nu:1.64\,, \vspace{2mm}\\
n:  -\,1.64\,;& e:\, 0.64\,.
\ea\right.
\ee

\noindent
$\lambda=5/(4\cos^2\theta)$  in (\ref{newq20}) decreases from 2 at the GUT scale to $\simeq 1.64$ at low energy, with $\sin^2\theta\simeq .238\,$.

\vspace{2mm}

This is in agreement with the conservation of $Q_U$ (following from those of $Q$ and $B-L$) by all interactions, strong, electroweak and grand-unified, in the massless $U$ limit,
with 
\be
\label{qlambda}
Q_U\,=\,\left\{\ba{ccc} W^\pm\  : &  \ \pm \,1 \,, \vspace{2mm}\\
Y^{\pm1/3}: &  \pm \,(Q_U^{\,u}+Q_U^{\,d})\! &= \,\pm \,\hbox{\small $\dis \frac{1-2\lambda}{3}$}\,, \vspace{2mm}\\
X^{\pm 4/3}: & \pm\,2\,Q_U^{\,u}&= \,\pm \,\hbox{\small $\dis \frac{4-2\lambda}{3}$}\,,
\ea
\right.
\ee
and

\vspace{-6mm}

\be
Q_U\,=\,\left\{\ba{cccccc} u: & \ \ \ \hbox{\small $\dis \frac{2-\lambda}{3}$}\,, 
&\ & p:& 1-\lambda\,,
\vspace{1.6mm}\\
d : & \ \   \hbox{\small $\dis \frac{-1-\lambda}{3}$}\,, 
&\ & n:& \ -\lambda\,,
\vspace{2.4mm}\\
\nu: &\ \ \  \ \ \lambda\,,\vspace{2.5mm}\\
e: & \ \ \ \lambda-1\,,
\ea
\right.
\ee
where $\lambda=5/(4\cos^2\theta)\simeq 1.64$ at low energy. The conservation of $Q_U$ by all interactions, including those induced by $X^{\pm 4/3}$ and $Y^{\pm 1/3}$ exchanges that could be responsible for proton decay, still unobserved \cite{pdec}, is illustrated in Fig.~\ref{fig:gut}.

\vspace{2mm}

Note that $\epsilon =\tan\chi\simeq (g'' /g')\cos\theta $ is already present at the grand-unification scale. As $C^\mu$  is decoupled from the visible sector for $\,g''\!=0\,$ (or $\tan\chi=0$) 
\footnote{Even if the notations suggest that $C^\mu$ is a free field for $g''=0\,$,  we can return to (\ref{lcur},\ref{f}) to still view as $C^\mu$ as possibly interacting with the hidden sector through the dark current ${\cal J}^\mu_{\,\rm d}$\,.},
 $\,\epsilon=\tan\chi$ may be viewed as a measure of the coupling between the visible and hidden sectors.
Furthermore a non-vanishing $\epsilon=\tan\chi$ cannot be generated just from quantum corrections once $\,g''\!=0\,$ so that $\epsilon$ vanishes at tree level, with $C^\mu$ interacting exclusively with the hidden sector, being decoupled from the visible one.

\section{\boldmath The $U$ \vspace{1.5mm}as a dark photon \hbox{\ \ \ \ \ \ also coupled to $B$ and $L$}}
\label{sec:5}

\vspace{-1mm}

\subsection{\vspace{1.2mm}Another orthogonal basis \hbox{\ \ \ \ \ \ for neutral gauge fields}}

\vspace{0mm} 

The $U$ current has been obtained  from the mixing (\ref{curr},\ref{curr2}) between the extra-$U(1)$ current $J^\mu_F$ and the standard weak neutral current $J^\mu_{\rm sm}$, providing, in the small $m_U$ limit, a vector current $J^\mu_U$ including a contribution 
$\propto J^\mu_{\rm em}$, with additional contributions from $B,\,L$ and dark matter currents.

\vspace{1.5mm}
This result may also be described in a complementary way, by constructing the same current  $J^\mu_U$ from the electromagnetic current $J^\mu_{\rm em}$ combined with the extra current $J^\mu_d$ in the hidden sector
(as in the specific ``dark photon'' case), but also in general  with the baryonic and leptonic currents 
$J^\mu_B$ and $J^\mu_{L_i}$.
Indeed the weak hypercharge current $J^\mu_Y$ in (\ref{cov},\ref{f}) may be viewed as coupled to $B^\mu$ and $C^\mu $ through the single hatted combination
\be
\label{zeta}
\hat B =\cos \zeta \,B + \sin\zeta \,C\,, \ \ \hbox{with}\  \tan \zeta= g''/g'\,.  
\ee

\vspace{2mm}
The doublet $\varphi$, with $F=Y=1$, 
\vspace{-.2mm}
interacts with $\hat B$ with the coupling constant $\,\hat g'=\sqrt{g'^2+g''^2}\,$. 
\vspace{-.5mm}
It does not interact with the orthogonal combination $\hat C$, which remains massless at this stage.
$\,<\!\varphi\!>\,$ generates  a spontaneous breaking of $\,SU(2)\times U(1)_{\hat Y} \to U(1)_{\hat{\rm QED}}$, leaving also $\hat A$ massless.
With
\be
\tan\hat \theta\,=\,\hat g'/g\,=\,\sqrt{g'^2\!+g''^2}/g\ 
\ee
this leads to define the orthonormal basis
\vspace{2mm}
\be
\label{33bis}
\framebox [8.5cm]{\rule[-2.4cm]{0cm}{5cm} $ \dis
\hspace{-.05mm}\hspace{1mm}\left\{\ \ba{ccccl}
\hat Z\!&=&\,\cos\hat \theta \ W_3-\sin\hat \theta\, \hat B \!&=&\!  \hbox{$\dis \frac{g \,W_3-g'B-g''C}{\sqrt{g^2+g'^2+g''^2}}$}
\vspace{2mm}\\ 
&&&\equiv &\ Z \ \ \hbox{in (\ref{33})}\ ,
\vspace{2.5mm}\\
\hat A\!&=& \,\sin\hat \theta \ W_3\!+\cos\hat\theta\, \hat B \!&=&\vspace{2mm}\\
&&&& \hspace{-22mm} =\,\hbox{$\dis \frac{(g'^2+g''^2)\,W_3+gg'B+gg''C}{\sqrt{g'^2+g''^2}\,\sqrt{g^2+g'^2+g''^2}}$}\,,
\vspace{3mm}\\
\hat C\!&=&\! -\sin \zeta \,B + \cos\zeta \,C \!&=&\!\hbox{$\dis \frac{-g''B+g'C}{\sqrt{g'^2+g''^2}}$}\,,
\ea\right.
$}
\ee
with $\hat \theta$ defined by
\be
\tan\hat \theta\,=\,\hat g'/g\,=\,\sqrt{g'^2\!+g''^2}/g\,.
\ee

\vspace{2mm}

We can also relate 
\vspace{-.2mm}
the two orthonormal basis in (\ref{33},\ref{33bis}), 
which have $Z=\hat Z$ in common, by writing, for the fields in the orthogonal plane,
\be
\label{rotchi}
\left\{\ba{ccr}
\hat A\!&=&\!\cos\chi\,A+\sin\chi\, U\,,
\vspace{2mm}\\
\hat C\!&=&\!-\sin\chi\,A+\cos\chi\, U\,.
\ea\right.
\ee
The angle $\chi$ between the $\hat A\,$ field in (\ref{33bis}) and the actual photon field $A$ in (\ref{33}) is obtained from the scalar products
\vspace{-3mm}
\be
\cos\chi= A. \hat A= U.\hat C\,,\ \ \sin\chi = -\,A.\hat C=\hat A.U\,,
\ee
so that
\vspace{1mm}
\be
\label{chi}
\left\{\ 
\ba{ccccc}
\cos\chi\!&=&\! g'\,\hbox{\small$\dis \sqrt{\frac{g^2\!+g'^2\!+g''^2}{(g^2+g'^2)(g'^2\!+g''^2)}}$} \!&=&\! \hbox{\small$\dis  \frac{g'\hat g_Z}{\hat g'g_Z}$}\ ,
\vspace{2mm}\\
\sin\chi\!&=&\! g g''/ \hbox{\small$\dis \sqrt{(g^2+g'^2)(g'^2\!+g''^2)}$}\!&=&\! \hbox{\small$\dis  \frac{g\, g''}{\hat g'g_Z}$} \ ,
\vspace{3mm}\\
\tan\chi\!&=&\! gg''/ g'\hbox{\small$\dis \sqrt{g^2\!+g'^2\!+g''^2}$}\!&=&\! \hbox{\small$\dis  \frac{g \,g''}{ g'\hat g_Z}$} \ ,
\ea \right.
\ee
giving back in particular (\ref{eps},\ref{eps2}) for $\epsilon=\tan\chi\,$.

\begin{figure}[t]\centering
    \includegraphics[scale=0.65]{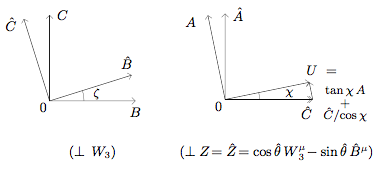}


\vspace{-1mm}

\caption{{\bf \em ``Mixing with the photon''},  
\vspace{0mm}
\em in the plane orthogonal to $Z$. 
\vspace{0mm}
 $\varphi$ interacts with $B^\mu\!$ and $C^\mu\!$ through $\,\hat B^\mu\!=$ $\cos\zeta \,B^\mu+\sin\zeta \,C^\mu$, leaving  $\,\hat C^\mu\!=-\sin\zeta \,B^\mu +\cos \zeta \,C^\mu$ massless.
$U^\mu$ is a (small) mixing of 
$\,\hat C^\mu$ 
(coupled to $B,\,L$ and dark matter) 
\vspace{-.3mm}
 \,with the photon field  $A^\mu$,
 \vspace{.3mm}
$U^\mu=(\hat C^\mu/\cos\chi) + \tan\chi\, A^\mu$,
\vspace{-.3mm}
 leading to $\,{\cal J}_U^\mu= \,({\cal J}^\mu_{\hat C}/\cos\chi )+e\tan\chi \,J^\mu_{\rm em}$ as in (\ref{ac2},\ref{curr4}), and earlier in (\ref{curr2}).
\vspace{-2mm}}
\label{fig:3}
\end{figure}

\subsection{The dark photon case}

In the specific dark photon case for which $F=Y+F_d$, 
standard model particles do not interact with $\hat C$, only with {\boldmath $\!W\!$} and $\hat B$ and thus $\hat Z$ and $\hat A\,$. 
In the small $m_U$ limit, $\hat A$
in (\ref{33bis}) appears for SM particles as a photon-like field, coupled to them only through their electromagnetic current, with strength 
  \be 
  \hat e\,=\,g\,\sin\hat \theta\,=\, g\ \hbox{\small$\dis \sqrt{\frac{g'^2\!+g''^2}{g^2\!+g'^2+g''^2}}$}=\,\hbox{$\dis  \frac{g \hat g'}{\hat g_Z}$}\,.
  \ee
The photon field, expressed as $
A= \cos\chi\,\hat A-\,\sin\chi \,\hat C\,$  where $\hat C$ is in this case uncoupled to SM particles, is coupled electromagnetically to them 
\vspace{-.3mm}
with the slightly reduced strength 
\be
\hat e \,\cos \chi = e=gg'/\sqrt{g^2\!+g'^2}\,.
\ee

\vspace{1mm}

At the same time  $U\!=\sin\chi \,\hat A+\cos\chi \,\hat C\,$ is coupled with the reduced strength $\hat e\,\sin\chi=e\,\tan\chi$.
The
$A$-$\hat A$ angle $\chi$, i.e.~in fact the $U$-$A$ mixing angle,  is obtained from 
   \be
 \!  \cos\chi=\frac{e}{\hat e}=g'\ \hbox{\small$\dis \sqrt{\frac{g^2\!+g'^2\!+g''^2}{(g^2\!+g'^2)(g'^2\!+g''^2)}}$}=\,\hbox{$\dis  \frac{g'\hat g_Z}{\hat g'g_Z}$} =
\,\hbox{$\dis  \frac{\sin\theta}{\sin \hat \theta}$}\,, 
   \ee
   as found in (\ref{chi}) in the general case.
$U$ is here coupled to SM particles only through their electromagnetic current, with strength $\,\hat e\, \sin\chi=e\,\tan\chi$, i.e.
\be
 \epsilon e= e\tan\chi =g''\hbox{\small $\dis \frac{g^2}{g^2+g'^2} \sqrt{\frac{g^2\!+g'^2}{g^2\!+g'^2\!+g''^2}}$} = g''\cos^2\theta\cos\xi, 
 \ee
 as directly found  in (\ref{eps}) from the mixing of  $J^\mu_F$ with  $J^\mu_{Z_{\rm sm}}$.
 
\subsection{\boldmath Recovering in a non-orthogonal \vspace{1.2mm}  basis \hbox{\ \ \ \ \ \,the general expression of the $U$ current}}

\vspace{-1mm}

The $\hat A/A/U$ mixing in (\ref{33bis}) provides an interpretation for the fact that the combination between the $J^\mu_F$ and $J^\mu_{Z_{\rm sm}}$ neutral currents provides a $U$ coupling to SM particles through the electromagnetic current.
But is it possible to extend this interpretation to the general situation involving also  $B$ and $L$ in the $U$ couplings\,?
\vspace{2mm}

To do so, 
\vspace{-.5mm}
we express from (\ref{rotchi}) the $U$ field in terms of the (non-orthogonal) $A$ and $\hat C$ fields defined in (\ref{33},\ref{33bis}), 
as  \vspace{-4mm}
\be
\label{ac}
U^\mu\,=\,\frac{\hat C^\mu}{\cos\chi}+ \tan\chi\, A^\mu\,.
\ee
Its equation of motion reads
\be
\label{ac2}
\framebox [7.2cm]{\rule[-.3cm]{0cm}{.8cm} $ \dis
\partial_\mu U^{\mu\nu}\,= \,{\cal J}_U^\nu\,=\,\frac{1}{\cos\chi}\ {\cal J}^\nu_{\hat C}\, + \,\tan\chi\ {\cal J}^\nu_{\rm em}\ .
$}
\ee

\noindent
With
\be
\label{currchat}
{\cal J}^\mu_{\hat C}=-\sin\zeta\,{\cal J}^\mu_Y +\cos\zeta\,{\cal J}^\mu_F =\cos\zeta\ \hbox{\small$\dis \frac{g''}{2}$}\,(\alpha J^\mu_B+\beta_i\,J^\mu_{L_i}+J^\mu_d),
\ee
the $U(1)$ gauge field $\hat C$ is now coupled, not just to the dark matter current as in the pure ``dark photon'' case, but  to a combination of it with the $B$ and $L$ currents.
\vspace{2mm}

Using the identity \footnote{This may be seen geometrically  from $\,\cos\zeta=  C.\, (\hat C=-\sin\chi \,A+\cos\chi \,U)=\cos\chi \,(C.U)=\cos\chi\cos\xi$\,, \,and verified using the expressions of $\cos\xi$, $\cos \zeta$ and $\cos \chi$
in (\ref{xi},\ref{zeta},\ref{chi}).}
\be
\label{ccc}
\frac{\cos\zeta}{\cos\chi}
=\,\hbox{\small $\dis \sqrt{\frac{g^2+g'^2}{g^2\!+g'^2\!+g''^2}}$}\,=\,\cos\xi\,,
\ee
and $g''\,\cos\xi\cos^2\theta=e\tan\chi $\,, we recover from (\ref{ac}-\ref{currchat})
expression (\ref{curr2}) of ${ \cal J}_U^\mu$, or equivalently
\be
\label{curr4}
{\cal J}_U^\mu\,=\,\epsilon\, e \,\left( J^\mu_{\rm em}+\frac{1}{2\,\cos^2\theta}\,(\alpha\,J^\mu_{B}+\beta_i\,J^\mu_{L_i}+J_d^\mu)\,\right).
\ee
This also reads, in a grand-unified theory,
\be
\label{curr5}
{\cal J}_U^\mu\,=\,\epsilon\, e \,\left( J^\mu_{\rm em}-\hbox{\small$\dis \frac{5}{4\,\cos^2\theta}$}\ J^\mu_{B-L}+\hbox{\small$\dis \frac{1}{2\,\cos^2\theta}$} \ J_d^\mu\,\right),
\ee
in agreement with expression (\ref{qu2}) of $Q_U$, with $\,\sin^2\theta\,\simeq \,.238\,$ as
 appropriate for low energies.
 
 \vspace{2mm}

The presence of the electromagnetic current, but also of the $B$ and $L$ currents, first found
from the mixing (\ref{22},\ref{curr}) between $J^\mu_F$ and $J^\mu_{Z_{\rm sm}}$ (cf.~Fig.\,\ref{fig:1}) \cite{plb89},  may be interpreted by expressing $U^\mu$ in terms of the non-orthogonal $A^\mu$  and $\hat C^\mu$ fields 
 as in (\ref{ac}) (cf.~Fig.\,\ref{fig:3}). ${\cal J}_U^\mu$ is then obtained in terms of $J^\mu_{\rm em}$ and $J^\mu_{d}$, $J^\mu_B$ and $J^\mu_{L_i}$ as in (\ref{ac2}-\ref{curr5}).
The specific case of a ``dark photon'' coupled proportionally to electric charges, simple to discuss and often used as a benchmark model, appears too restrictive, and the
possible couplings of the $U$  to the $B$ and $L$ currents should be taken into account as well.

\subsection{``Kinetic mixing'' as the effect \vspace{1mm} of a description in a non-orthogonal field basis}

As a side remark,  the notion of ``kinetic mixing'', popular now, has been used nowhere. It goes without saying that, 
as in {\em any\,} theory (including the standard model itself), using a non-orthogonal rather than an orthonormal field basis would {\em introduce} in the expression of the Lagrangian density non-diagonal terms, now often referred to\, as ``kinetic-mixing'' terms. These can always be eliminated by returning to an orthonormal basis, without any loss of physical content. Furthermore, and in contrast with a general belief, {\it it is not necessary that these ``kinetic-mixing'' terms be gauge invariant}, provided of course the complete Lagrangian density is invariant, independently of the basis in which it is expressed.

\vspace{2mm}

To illustrate this
let us rewrite the Lagrangian density defined in terms of $B^\mu$ and $C^\mu$ as in (\ref{cov}-\ref{f}), 
\vspace{-.5mm}
 in the non-orthogonal basis
 ($\hat B^\mu,\,C^\mu$), or ($B^\mu,\hat C^\mu$), represented in Fig.~\ref{fig:3},  using from (\ref{zeta},\ref{33bis}) 
 \be
 B^\mu=(\hat B^\mu /\cos\zeta)-\tan\zeta \,C^\mu\,.
 \ee 
We get
\be
\label{lk}
 \hspace{-1.5mm}\ba{ccl}
{\cal L}_k\!\!&=&\!-\,\hbox{\small$\dis \frac{1}{4}$}\, \left( W^{\mu\nu}_{\,3}\,W_{3\,\mu\nu}+ B^{\mu\nu}B_{\mu\nu}+ C^{\mu\nu}C_{\mu\nu}\right)
\vspace{1.5mm}\\
&=&\!-\,\hbox{\small$\dis \frac{1}{4}$}\ W^{\mu\nu}_{\,3}\,W_{3\,\mu\nu}
\vspace{1mm}\\
\!&&\!   \ \  -\,\hbox{\small$\dis \frac{1}{4} \,\frac{1}{\cos^2\zeta}$}\left( \hat B^{\mu\nu}\hat B_{\mu\nu}\!+ C^{\mu\nu}C_{\mu\nu}\!- 2\sin\zeta\, \hat B^{\mu\nu}C_{\mu\nu}\right)
\vspace{2.5mm}\\
&=&\!-\,\hbox{\small$\dis \frac{1}{4}$}\ W^{\mu\nu}_{\,3}\,W_{3\,\mu\nu}
\vspace{1mm}\\
\!&&\!    \ \ -\,\hbox{\small$\dis \frac{1}{4} \,\frac{1}{\cos^2\zeta}$}\left( B^{\mu\nu}B_{\mu\nu}\!+ \hat C^{\mu\nu}\hat C_{\mu\nu}\!+2\sin\zeta\, B^{\mu\nu}\hat C_{\mu\nu}\right)\!.
\ea 
\ee
This is immediately rediagonalized 
by returning to the orthogonal fields $B^\mu$ and $C^\mu$ (or $\hat B^\mu$ and $\hat C^\mu$).

\vspace{2mm}

{\em The same\,} kinetic terms (\ref{lk}) in $\cal L$ may be reexpressed  in other non-orthogonal basis, involving 
different non-diag\-onal ``kinetic-mixing'' terms, this time not even gauge invariant. With
\be
\label{kmix}
\ba{rcc}
U^\mu\!&=&\hbox{\small$\dis \frac{\hat C^\mu}{\cos\chi}$}\,+ \,\tan\chi\ A^\mu\,,
\vspace{2mm}\\
\hbox{or}\ \ \ \ A^\mu\!&=&\hbox{\small$\dis \frac{\hat A^\mu}{\cos\chi}$}\, - \tan\chi\ U^\mu\,,
\ea
\ee
as in (\ref{ac}) one has
\be
\label{lka}
\ba{ccl}
\hspace{-1mm}{\cal L}_k\!&=&\!-\,\hbox{\small$\dis \frac{1}{4}$}\, \left( Z^{\mu\nu}\,Z_{\mu\nu}+ A^{\mu\nu}A_{\mu\nu}+ U^{\mu\nu}\,U_{\mu\nu}\right)
\vspace{1.5mm}\\
\!&=&\!-\,\hbox{\small$\dis \frac{1}{4}$}\ Z^{\mu\nu}\,Z_{\mu\nu}
\vspace{1mm}\\
\!&&\!\  -\,\hbox{\small$\dis \frac{1}{4} \,\frac{1}{\cos^2\chi}$}\left(A^{\mu\nu}A_{\mu\nu}\!+ \hat C^{\mu\nu}\hat C_{\mu\nu}\!+2\sin\chi\, A^{\mu\nu}\hat C_{\mu\nu}\right)\!,
\vspace{2.5mm}\\
\!&=&\!-\,\hbox{\small$\dis \frac{1}{4}$}\ Z^{\mu\nu}\,Z_{\mu\nu}
\vspace{1mm}\\
\!&&\!\ -\,\hbox{\small$\dis \frac{1}{4} \,\frac{1}{\cos^2\chi}$}\left(\hat A^{\mu\nu}\hat A_{\mu\nu}\!+ U^{\mu\nu}U_{\mu\nu}\!-2\sin\chi\, \hat A^{\mu\nu}U_{\mu\nu}\right)\!,
\ea
\ee
where  the ``kinetic-mixing'' terms, which no longer involve abelian gauge fields only,  are not gauge invariant.

\vspace{2mm}

The mixing angles between $\hat C^\mu$ and 
$A^\mu$ in (\ref{lka}), or $\hat C^\mu$ and  $B^\mu$ 
 in (\ref{lk}) are geometrically related by
  \be
 \sin\chi =-\,\hat C.\,A=\cos\theta \, (\,\underbrace{-\,\hat C.B}_{\hbox{\small $\sin\zeta$}}\,)-\,\sin\theta \ \underbrace{\hat C.\,W_3}_{\hbox{\small$0$}}\,,
 \ee
 
 \vspace{-2mm}
 
 \noindent
 thanks to the orthogonality between $\hat C$ and $W_3$ (cf.\,Fig.\ref{fig:3}),
 i.e.
 \vspace{-2mm}
 \be
 \sin\chi= \sin\zeta \cos\theta\,,
 \ee
easily verified from  $\sin\zeta\cos\theta= (g''/\hat g') \,(g/g_Z)= \sin\chi\,$.

\vspace{2mm}

Altogether there is no real gain in considering such non-diagonal kinetic terms, immediately rediagonalized by returning to the original expressions. 
Considering ki\-netic-mixing terms i.e.~using non-orthogonal field basis simply appears as a substitute for the introduction of the appropriate couplings of $C^\mu$ 
in the covariant derivative  (\ref{cov}), involving both the visible and hidden sectors and  leading to the corresponding current ${\cal J}^\mu_F$ in (\ref{f}). This should not hide 
that $C^\mu$ may be coupled to $B$ and $L$ as well as to $Y$ and dark matter, and the $U$ boson to a combination of the electromagnetic with the $B,\,L$ and dark matter currents. 
\vspace{2mm}

In addition, a non-vanishing mixing angle $\chi$, and coupling $\epsilon =e\,\tan\chi$ 
\vspace{-.5mm}
(also relating the visible and hidden sectors),  corresponding to $U^\mu\!= (\hat C^\mu/\cos\chi)+\tan\chi\,A^\mu$ as in (\ref{ac}), may be obtained directly even in the presence of a single $U(1)$ gauge group, as in a $SU(5)\times U(1)_F$ gauge theory, with  the visible and hidden sectors getting totally decoupled for $\,\epsilon=\chi=\zeta=\xi=0$ \footnote{There is thus no need to invoke hypothetical effects of radiative corrections for generating a non-vanishing value of $\epsilon$, especially when no coupling between the visible and hidden sectors is present.}.
These non-vanishing $\epsilon$ and $\chi$ are obtained here in spite of the  non-existence of a gauge-in\-variant kinetic-mixing term between non-abelian ($SU(5)$) and abelian ($U(1)_F$) gauge fields.

\section{\boldmath Implications for a light $U$}
\label{sec:6}

\vspace{-.5mm}

\subsection{Axial couplings are strongly constrained}
\label{subsec:6a}
\vspace{-2mm}

The axial couplings of the $U$ should satisfy
  \be
     \label{pv}
     (f_e^A\,f_q^V)^{\frac{1}{2}}\ \simle \ 10^{-7}\ m_U(\hbox{MeV})\,,
     \ee
for $m_U$ larger than a few MeVs, expressing that $|f_e^A\,f_q^V|/m_U^2\simle 10^{-3}\, G_F\,$, to avoid too-large parity-viol\-ation effects in atomic physics \cite{plb05}.
\ghost{
Axial and vector parts in the $U$ current could lead to too large parity-violation effects in atomic physics, providing, from
$|f_e^A\,f_q^V|/m_U^2\,\simle 10^{-3}\,G_F$, 
 the strong constraint \cite{plb05}
     \be
     \label{pv}
     (f_e^A\,f_q^V)^{\frac{1}{2}}\ \simle \ 10^{-7}\ m_U(\hbox{MeV})\,.
     \ee
     }
     \hspace{-3mm}
A light $U$ with axial couplings ($1^+$) could also be produced in a longitudinal polarization state with enhanced effective pseudoscalar couplings to quarks and leptons \cite{npb81},
\be
f_{ql}^P=\,f_{ql}^A\ \ \frac{2\,m_{ql}}{m_U}\,,
\ee
much like a $0^+$ pseudoscalar $a$.  The resulting constraints on the axial couplings to heavy quarks, from $\psi$ or $\Upsilon\to\gamma \,U $ and $K^+\to\gamma \,U $ decays \cite{plb86}, are rather severe, especially for a light $U$ with invisible decay modes  into $\nu\bar \nu$ or LDM particles. In particular the axial couplings
to down quarks and charged leptons, universal when they get  masses from the same doublet v.e.v. (as in supersymmetric theories)
must then verify 
\cite{prd06,prd07,plb09}
\be
     \label{fa}
f_{e,d}^A \simle\ (2\ \hbox{to}\ 4)\ 10^{-7} \,m_U\hbox{(MeV)}/\sqrt{B_{\rm inv}}
\ee
(which is typically $\simle 10^{-5}$  for a $U$ in the $\sim 10$ MeV mass range). 
This leads, in such cases where axial couplings may occur, to consider a $U(1)_F$ symmetry broken at a scale larger than electroweak through a large singlet v.e.v., very much as for an ``invisible''  axion \cite{plb80}. 

\vspace{2mm}

The $U$ lifetime may vary considerably between less than $10^{-15}$ s to many years and even infinity, depending on its mass and 
couplings. The  decay rate for $U\to e^+e^-$ is given by \cite{npb81}
\be
\ba{ccc}
\Gamma_{ee}\!&\simeq&\! \! \dis \frac{1}{12\pi}\ \left[\,(f_e^V)^2\,  \left(1+\frac{2m_e^2}{m_U^2}\right)\,\sqrt {1-\frac{4m_e^2}{m_U^2}}\,\right.\ \ \ \ 
\vspace{2mm}\\
&&\hspace{15mm}  +\, \dis  \left.\,(f_e^A)^2 \, \left(1-\frac{4m_e^2}{m_U^2}\right)^{\!3/2}\ \right] m_U \,,
\ea
\ee
including  the phase-space factors 
\vspace{-.6mm} 
$\frac{3}{2}\,\beta-\frac{1}{2}\,\beta^3$ and $\beta^3$ for the vector and axial production of  spin-$\frac{1}{2}$ particles,
\vspace{-.6mm} 
with $\beta =\!\sqrt{1-4m_e^2/m_U^2}\,$.
We also have to take into account 
the invisible decay modes of the $U$ into ordinary neutrinos, given for three left-handed neutrinos with chiral couplings $f_\nu$ by

\vspace{-5mm}
\be
\Gamma_{\nu_L\,\overline{\nu_L\!\!}} \  \simeq \, \frac{f_\nu^2}{8\pi} \ m_U\,,
\ee
and other possible invisible decays into right-handed neutrinos and light dark matter particles, which could decrease significantly the branching ratio for  $U\to e^+e^-$.

\vspace{-1mm}

\subsection{Vector couplings should not be too large}

\vspace{-1mm}

Let us thus return to a light $U$ vectorially coupled to SM particles, as is the case when a single doublet $\varphi$ contributes to the electroweak breaking (or several but with the same gauge quantum numbers) \cite{plb89}.
Couplings proportional to a combination of $B,\,L_i$ and  $Q$ as in (\ref{qu},\ref{newq20},\ref{curr5}), rather than just $Q$, open new possibilities for experimental detection \cite{prd06}. 
Experimental results, usually discussed in the $(\log \epsilon, m_U)$ plane, should
also be considered in terms of these couplings $\,f\!=\epsilon e\,Q_U$,
 through the changes in the couplings to SM particles
\be
\label{newq4}
\epsilon e\,Q\ \to \ f\,=\,\epsilon e\,Q_U =\, \epsilon e\ (Q+\lambda_B B+\lambda_{L_i} L_i)\,,
\ee
leading to 
$
f_n+f_{\nu_e} = f_p+f_e
$ as in  (\ref{rel2}),
and more specifically
\be
\label{qlambda2}
\epsilon e\,Q\ \to \ \epsilon e \,Q_U = \epsilon e \ (Q- \lambda (B-L))\,,
\ee
with $Q_U(e)=\lambda-1\simeq .64$ in a grand-unified theory.
This usually results in moderate shifts of the various limits when expressed in terms of $\log \epsilon$\,; e.g.~for an experiment sensitive to the couplings to the electron, through the change
\be
(\epsilon e)^2_{\hbox{\footnotesize \ dark photon}}\  \to\ [\,f_e=\epsilon e\,Q_U(e)\,]^2 \times B_{ee}\,.
\ee

\vspace{2mm}

The $U$ boson should interact sufficiently weakly with electrons, so that  its
contribution to the electron anomaly, 
\vspace{-3mm}
\be
\delta_U a_e\,\simeq \,\frac{f_e^2}{12\pi^2}\ \frac{m_e^2}{m_U^2}\simeq\,\frac{\alpha}{3\pi}\ [\,\epsilon \,Q_U(e)\,]^2\,\frac{m_e^2}{m_U^2}
\ee
 for  $m_U$ larger than a few MeVs, be less than about $3\ 10^{-12}$ \cite{ge-2} (improved over the earlier $2\ 10^{-11}$ leading to
$\,|f_e|\,\simle\, 10^{-4}\ m_U$(MeV) \cite{prd07}). This requires
\be
\label{fe}
\left|\,f_e= \epsilon e \,Q_U(e)^{\phantom 2}\!\!\right| \  \simle \ 4\ 10^{-5}\,m_U(\hbox{MeV})\,,
\ee
or
\be
\label{fe2}
\left|\,\epsilon\,Q_U(e)^{\phantom 2}\!\!\right| \  \simle \ 1.2\ 10^{-3}\,\frac{m_U}{10\ \hbox{MeV}}\,,
\ee
applicable for $m_U$ above  a few MeVs.
\vspace{2mm}

The $U$ should also act sufficiently weakly with protons, so that the  $\pi^0 \to \gamma \,U$ decay rate be sufficiently small.
 The corresponding branching ratio  is given, for $m_U$ somewhat below $m_{\pi^\circ}$, by
$\,2\,(2f_u+f_d)^2/e^2 =2\, \epsilon^2 \,Q_U(p)^2$, 
replacing the $2\,\epsilon^2$ of the pure dark photon case, typically constrained to be $\,\simle 10^{-6}$ for $m_U$ in the 10-100 MeV range \cite{NA48}. This experiment provides similar limits for a $U$ boson decaying into $e^+e^-$, with the replacement
\be
(\epsilon e)^2_{\hbox{\footnotesize \ dark photon}}\  \to\ [\ \underbrace{f_p=2f_u+f_d}_{\hbox{\small$\epsilon e\,Q_U(p)$}}\ ]^2 \times B_{ee}\,,
\ee

\vspace{-5mm}

\noindent
leading to
\be
\label{fp}
\left|\,f_p=\epsilon e\,Q_U(p)^{\phantom 2}\!\right|  \ \simle \ 3\ 10^{-4}\,/\,\sqrt {B_{ee}}\,,
\ee
with $Q_U(p) = 1-\lambda\simeq\,-\,.64$ in a grand-unified theory.

\vspace{2mm}
For a boson with significant invisible decays 
into neutrinos or light dark matter particles, we have,
from the search for the decay $\pi^0\to \gamma +U_{\rm \, inv.}$ 
 \cite{uinv,uinv2,nomad}, with a branching ratio fixed by $\,2\epsilon^2\, Q_U(p)^2 \!< \,3.3\ 10^{-5}$
for $m_U< $ 120 MeV, the limit 
\be
\label{fp2}
\left|\,f_p=\epsilon e\,Q_U(p)^{\phantom 2}\!\right|  \ \simle \ 1.2\ 10^{-3}\,/\,\sqrt {B_{\rm inv}}\,.
\ee

\vspace{1mm}

We also have, from  a low-$q^2\ \nu_e$-$e$ scattering experiment \cite{lsnd}, the constraint $\,|f_{\nu_e} f_e|/m_U^2$ $\simle G_F$ \cite{npb04,plb05},
\,i.e.  
\be
\label{nue}
|f_{\nu_e} f_e|^{1/2} \,\simle \,\, 3\ 10^{-6} \ m_U(\hbox{MeV})\,,
\ee
valid for  $m_U$  larger than a few MeVs (or $\,|f_{\nu_e} f_e|/m_U^2 \simle 10^{-5}$ otherwise), also expressed as
\be
\label{nue2}
\epsilon \ |\,Q_U(\nu_e)\,Q_U( e)\,|^{1/2} \ \simle \,\, 10^{-4} \ \frac{m_U}{10 \ \hbox{MeV}}\ .
\ee
If $f_e, f_p$ and $f_\nu$ are all small as suggested by the above constraints, $f_n$ may have to be small as well, as a result of (\ref{rel2}).

 \vspace{2mm}

For an anomaly-free theory the currents may be constructed from $Q$, $B-L$ (with $\nu_R$'s) and $L_i-L_j$, with the dark matter current, also vectorial, involving spin-0 or Dirac \hbox{spin-1/2} dark matter particles. With a family-independent symmetry  $Q_U$ involves $B\!-\!L$ as in (\ref{newq20-1}), $\,Q_U\! = Q$ $-\,\lambda\,(B-L)$ $ +\,Q_{U\,\rm dark}$,
as  found in a grand-unified theory, implying
\be
f_e =-\,f_p= (\lambda-1)\,\epsilon e\,,\ f_\nu=-f_n=\lambda\, \epsilon e\,,
\ee
in agreement with (\ref{rel2}). More specifically 
the constraint (\ref{nue},\ref{nue2}) reads $\,\epsilon e \,\sqrt{\left|\lambda (\lambda-1)\right|} \simle \,\,3\ 10^{-5} \ m_U/(10$ MeV)\, and typically  implies, in a grand-unified theory with $\lambda\simeq 1.64$ so that $ \sqrt{\left|\lambda (\lambda-1)\right|} \simeq 1$,
\be
\epsilon  \ \simle \, 10^{-4} \ \frac{m_U}{10 \ \hbox{MeV}}\ ,
\ee
for $m_U$ larger than a few MeVs.

\vspace{2mm}

Still it may be possible, although at the price of elegance, 
to arrange for $\nu_e$, or $\nu_e$ and $\nu_\mu$, not to interact with $U$, e.g.~through the change
$\,B\!-\!L\to 
B-3L_\tau$, leading to  
\vspace{-2mm}
\be
Q_U = \,Q-\lambda\,(B-3L_\tau) +Q_{U\,\rm dark}\,,
\ee
still in an anomaly-free theory. More specifically 
\be
Q_U\simeq \,Q-(B-3L_\tau)+Q_{U\,\rm dark}\,
\ee

\noindent
would lead to $\,Q_U(n)\simeq Q_U(e) \simeq Q_U(\mu) \simeq -1$, $Q_U(p)\simeq $ $Q_U(\nu_e)\simeq Q_U(\nu_\mu)\simeq 0$, i.e.
\be
f_n\simeq f_e\simeq\,-\,\epsilon \,e\,,\  \ \hbox{with } \ f_p,\,f_{\nu_e},f_{\nu_\mu}\ \ \hbox{very small\,.}
\ee

\vspace{1mm}

The $U$ couplings in (\ref{qu}) offer new opportunities for tentative interpretations of possible anomalies.
The decays of excited states of $^8$Be have long been viewed as potentially sensitive to the production of light \hbox{spin-1} $U$ bosons and anomalous production of $e^+e^-$ pairs that could signal such decays have already been reported \cite{vit}, although  these first indications were not confirmed. A possible new anomaly 
has been found recently \cite{k16}, which remains to be better understood 
before attempting at an interpretation \cite{feng, feng2}.
It does not seem that it can 
\linebreak
 be attributed to a dark photon coupled proportionally to electric charges,
which would require a too large  $\epsilon$.
Expressions (\ref{curr2},\ref{qu}) of the $U$ charge and current \cite{plb89}
may help provide an interpretation if the effect is real, possibly with a $U$ interacting more strongly with neutrons and electrons than with protons and  neutrinos, keeping in mind relations (\ref{rel}-\ref{rel2}) associated with a conserved $Q_U$ in the massless limit.

\vspace{7mm}

\section{Conclusions}

\vspace{3mm}

\noindent
{\em Expression of $\,Q_U$ from $Q,\ B$ and $L$}

\vspace{2.5mm}

The structure presented here for the interactions of a light $U$ boson depends on a very small number of relevant parameters, especially   $\epsilon$ and $m_U$ with $Q_U$ expressed as $Q+\lambda_B B+\lambda_i\,L_i\ [+\,Q_d\,]\,$ or $\,Q-\,\lambda \,(B-L)\ [+\,Q_d\,]\,$.
\linebreak
It provides a consistent framework to deal with  a new interaction, naturally parity-conserving in the visible sector and coupled to a conserved charge
$Q_U$, in the small mass limit.  The $U$ may be viewed as {\it mixed with the $Z$}, or {\it mixed with the photon}, or both at the same time with a $\,3\times 3\,$ mixing matrix extending the $\,2\times 2\,$ electroweak one of the standard model.

\vspace{5.5mm}

\noindent
{\em Kinetic mixing  as an effect of a non-orthogonal field basis}

\vspace{3mm}

What is often referred to as ``kinetic mixing'' simply corresponds to choosing a description in a non-orthogonal field basis. This also implies  that the  kinetic-mixing terms associated with this basis are not even required to be gauge invariant, in contrast with a common belief. 
Furthermore, even with the $U$ viewed as kinematically mixed with the photon, 
$B$ and $L $ contributions are generally allowed  in its couplings, and  may even be required as in the case of grand-unification.

\vspace{2mm}

\pagebreak

\vbox{
\noindent
{\em A non-vanishing $\,\epsilon=\tan\chi \,$ within grand-unification}

\vspace{2.5mm}
The construction is compatible with grand-uni\-fication,  with  a charge $Q_U$ commuting with the electrostrong symmetry between the photon and gluons at the GUT scale, and  {\it a non-vanishing $\,\epsilon=\tan\chi\,$ already present at the classical level.} 
This occurs in spite of the fact that the $SU(5)\times U(1)_F$ gauge group includes 
\vspace{-.2mm}
{\it a single abelian factor} \ $U(1)_F$, so that no gauge-invariant kinetic term mixing the $SU(5)$ and $U(1)_F$ gauge fields may be written.
Still it is possible to view the $U$ as (``kinematically'') mixed with the photon as in (\ref{ac},\ref{kmix}), with non gauge-invariant mixing terms in the lagrangian density as in (\ref{lka}), and a $U$ boson also coupled to the $B$ 	and $L$ currents.
}

\vspace{4.5mm}
\vbox{
\noindent
{\em $Q_U$ from the $\,SU(4)_{\rm es }\!\times U(1)_U$ electro\-strong symmetry}

\vspace{2.5mm}

$Q_U$ evolves, from  $\,Q-2\,(B-L) + Q_{U\,\rm dark}$ at the GUT scale, to
\vspace{-2mm}
\be
Q_U\,\simeq \ Q-1.64 \,(B-L) + Q_{U\,\rm dark}
\ee

\noindent
 at low energy.
This expression, and the more general one $\,Q_U\simeq \,Q-\lambda \,(B-L) + Q_{U\,\rm dark}$, motivated by grand-unification and by anomaly-cancellation,  may be used to {\it display the experimental constraints in the $(\log \epsilon,\,m_U)$ plane as a function of $\,\lambda\,$.}
}
\vspace{2.5mm}

The $U$ current in the visible sector is purely vectorial in the massless $U$ limit, in relation with the fact that the theory may admit at high energy 
{\it an extended  $\,U(4)= SU(4)_{\rm es}\times U(1)_U$ electrostrong symmetry which preserves parity,}\,
with the $\,U$, \,photon, gluons and $X^{\pm4/3}$ bosons all coupled to vector currents. The  interactions mediated by the $Z$ and the electrostrong quartet $(Y^{\mp1/3}\!, \, W^\pm)$, on the other hand, violate parity.

\vspace{-1.5mm}
\bc
*\hspace{4mm}*
\ec
\vspace{-3mm}

A large variety of interesting effects may occur, in particular for a  $U$ in the $\approx$ MeV to hundred MeVs mass range.  
The $U$ boson, if extremely light or massless, may also lead to a new long range force, extremely weak,  that could manifest through apparent violations of the Equivalence Principle. This one will soon be tested in space to an increased level of precision.
The characteristics of the new interaction mediated by such a light neutral \hbox{spin-1} $U$ boson may also play a role in shedding light on a possible unification of weak, electromagnetic and strong interactions.

\bibliography{References}

\end{document}